\documentclass[conference]{IEEEtran}
\IEEEoverridecommandlockouts
\usepackage{cite}
\usepackage{amsmath,amssymb,amsfonts}
\usepackage{algorithmic}
\usepackage{graphicx}
\usepackage{textcomp}
\usepackage{xcolor}
\usepackage{tikz}
\usetikzlibrary{tikzmark}
\usetikzlibrary{shapes,arrows,positioning}
\usepackage{listings}
\usepackage{multirow}
\usepackage{pifont}
\newcommand{\cmark}{\ding{51}}%
\newcommand{\xmark}{\ding{55}}%

\usepackage{soul}

\usepackage[hidelinks]{hyperref}
\usepackage{cleveref}

\usepackage{enumitem}

\makeatletter
\let\MYcaption\@makecaption
\makeatother
\usepackage[font=footnotesize]{subcaption}
\makeatletter
\let\@makecaption\MYcaption
\makeatother

\def\BibTeX{{\rm B\kern-.05em{\sc i\kern-.025em b}\kern-.08em
    T\kern-.1667em\lower.7ex\hbox{E}\kern-.125emX}}

\input{sec/00preamblelistingsconf}

\begin{document}

\title{Evaluating LLMs for Hardware Design and Test}


\author{\IEEEauthorblockN{Jason Blocklove}
\IEEEauthorblockA{\textit{New York University}\\
New York, NY USA \\
jason.blocklove@nyu.edu}
\and
\IEEEauthorblockN{Siddharth Garg}
\IEEEauthorblockA{\textit{New York University}\\
New York, NY USA \\
siddharth.garg@nyu.edu}
\and
\IEEEauthorblockN{Ramesh Karri}
\IEEEauthorblockA{\textit{New York University}\\
New York, NY USA \\
rkarri@nyu.edu}
\and
\IEEEauthorblockN{Hammond Pearce}
\IEEEauthorblockA{\textit{University of New South Wales} \\
Sydney, Australia \\
hammond.pearce@unsw.edu.au}
}


\maketitle

\begin{abstract}
Large Language Models (LLMs) have demonstrated capabilities for producing code in Hardware Description Languages (HDLs). However, most of the focus remains on their abilities to write functional code, not test code. The hardware design process consists of both design and test, and so eschewing validation and verification leaves considerable potential benefit unexplored, given that a design and test framework may allow for progress towards full automation of the digital design pipeline.
In this work, we perform one of the first studies exploring how a LLM can both design and test hardware modules from provided specifications. Using a suite of 8 representative benchmarks, we examined the capabilities and limitations of the state-of-the-art conversational LLMs when producing Verilog for functional and verification purposes. We taped out the benchmarks on a Skywater 130nm shuttle and received the functional chip.
\end{abstract}

\begin{IEEEkeywords}
Hardware Design and Verification, CAD, LLM
\end{IEEEkeywords}

\section{Introduction}

Digital hardware design has traditionally relied on a relatively niche skill set, requiring specifically trained engineers to both architect and implement new semiconductor hardware designs, and then to validate and verify those designs before they are taped out into integrated circuits. 
This testing process is time-consuming and difficult, requiring the development of appropriately comprehensive tests to ensure that every possible eventuality is accurately accounted for. Indeed, according to \cite{foster_wilson_2020}, 51\% of development effort (cost) in both ASIC and FPGA-based systems are spent on verification. 
Any missed bugs will have increasingly expensive consequences, depending on how late in the product's development and deployment lifecycle they are eventually discovered.
In a push to simplify design and test, tools have been developed to support processes like high-level synthesis (HLS)~\cite{coussy_high-level_2010}, which make it easier for those with preexisting software development experience to create functioning hardware---though in order for an engineer to take full advantages of the domain, they will still often require a good level of hardware development knowledge.

Other techniques have leveraged machine learning (ML) to enhance  tooling for design and test. For design-focused examples~\cite{hamolia_survey_2021} discusses ML adoption in logic synthesis, design space reduction, exploration, placement, and routing. These techniques focus on simplifying  electronic design automation (EDA) algorithms rather than simplifying the  tasks. 
Advances in natural language processing---in particular, Large Language Models (LLM)---have presented new avenues for hardware design. 
Prior work shows that LLMs can write code, including in verilog hardware description language (HDL)  ~\cite{pearce_dave_2020,thakur_benchmarking_2023,thakur_verigen_2024,blocklove_chip-chat_2023,liu_verilogeval_2023}. However, while this demonstrated how LLMs can aid \textit{design}, research exploring LLMs for \textit{test} are lacking.




In this work we evaluate how LLMs may be used for \textit{both} design and test \textit{together} by 
starting from a plain language specification, instructing for Verilog implementation, and then creating the associated testbenches. 
To this end we leverage design and test tools in a feedback loop with a LLM such that minimal user debugging assistance is required. 
To this end, we provide a set of conversational LLMs with simple prompts to create the functional and verification HDL, which we then successfully taped out through Tiny Tapeout 3~\cite{noauthor_tiny_2023}.
Our contributions are:
(i) Developing simple benchmarks to evaluate the capabilities of LLMs for functional hardware development and verification.
(ii) Providing post-silicon results from a taped-out device.
\textbf{Open-source:} Benchmarks, toolchain scripts, Verilog and LLM conversation logs are on Zenodo~\cite{review_repository_2024}.

\section{Background and Related Work}
\label{sec:background}

\subsection{Large Language Models (LLMs)}
\label{sec:llms}
LLMs such as GPT-3~\cite{brown_language_2020} and Codex~\cite{chen_evaluating_2021} are a class of ML model which use a Transformer architecture~\cite{vaswani_attention_2017} and are trained on a large corpus of plain-language data to generate a predicted output.
LLMs receive an text-based input prompt, then produce the ``most-likely'' continuation of that prompt. 
By training over appropriate data this means LLMs can complete lexical sequences, including code.
Further, LLMs have recently been made ``conversational'' using instruction-tuning.
Rather than guessing the next most likely token in an ``autocomplete'' fashion, they ingest whole prompts and formulate complete responses to those prompts. Examples  include ChatGPT~\cite{openai_chatgpt_2022} versions 3.5 and 4, Google's Bard~\cite{pichai_important_2023}, and HuggingFace's HuggingChat~\cite{huggingface_huggingchat_2023}.\

\subsection{LLM Aided Design}
Several LLMs have been created with the intention of generating Verilog for hardware design, e.g. Pearce et al.~\cite{pearce_dave_2020} fine-tuned a GPT-2 model aimed at generating Verilog; and VeriGen~\cite{thakur_benchmarking_2023} is a fine-tuned CodeGen model; another work just used GPT-4 directly to co-author a microprocessor~\cite{blocklove_chip-chat_2023} sent for tapeout. 
Commercial models also exist, such as RapidGPT~\cite{noauthor_welcome_nodate}, ChipNeMo~\cite{liu_chipnemo_2023},  JedAI~\cite{noauthor_cadence_nodate}, and Synopsys.ai~\cite{noauthor_redefining_nodate}.
Test-focused work is typically aimed at evaluating the models, e.g. VerilogEval~\cite{liu_verilogeval_2023} offers a set of standard benchmarks derived from HDLBits~\cite{noauthor_problem_nodate} for LLMs writing Verilog.
LLM work focusing on \textit{design} test is more nascent. \cite{kande_security_2024} explored generating SystemVerilog assertions with some success, and \cite{ahmad_flag_2023} examined bug-finding in isolation without considering tool feedback.

\section{Prompting LLMs for Design and Test}
\label{sec:scripted}


\subsection{Methodology}
\label{sec:conversation-flow}
\begin{figure}[t]
    \centering
    \includegraphics[width=\columnwidth]{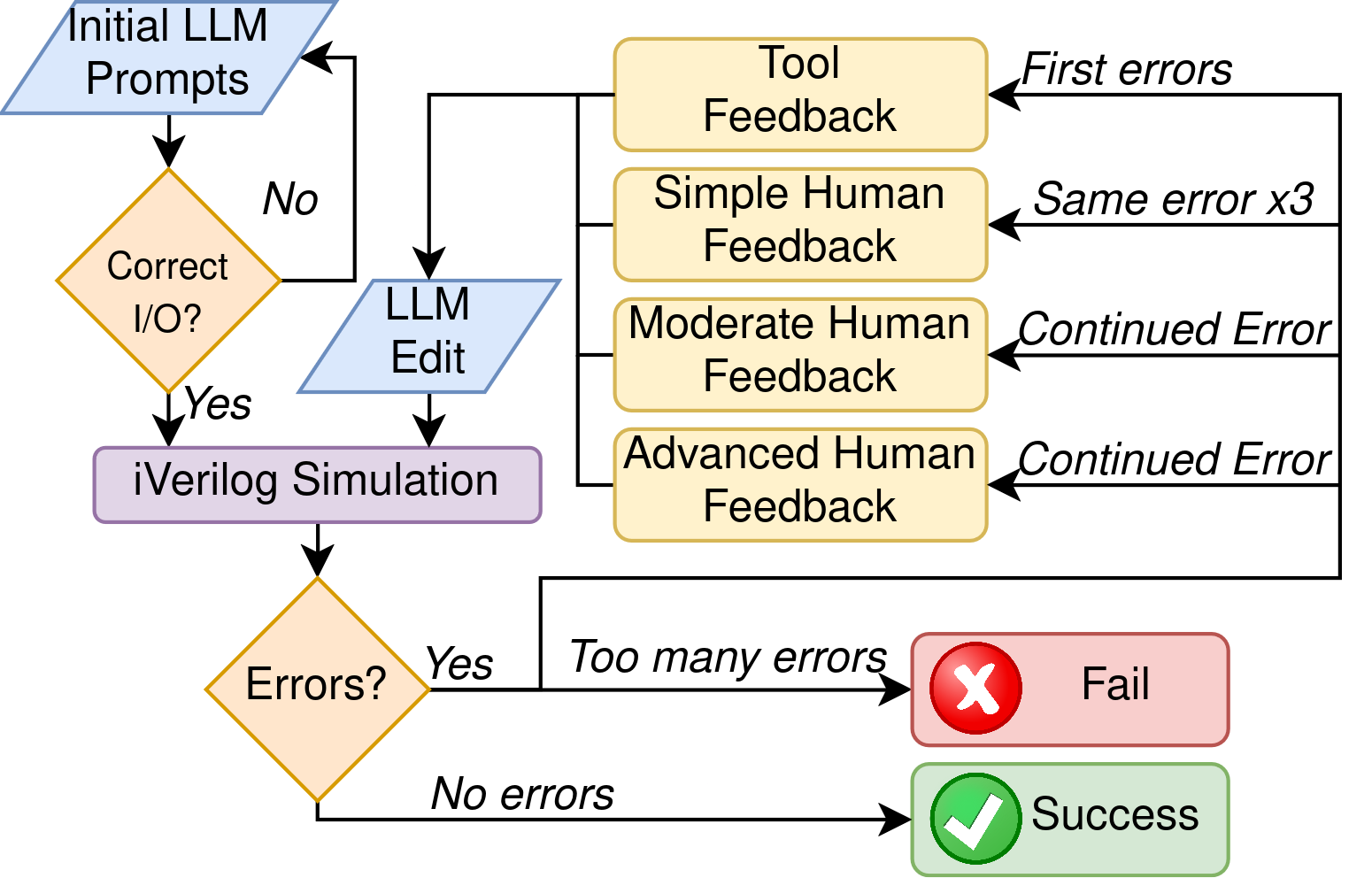}
    \caption{Simplified LLM conversation flowchart}
    \label{fig:conversation_flow}
\end{figure}

\Cref{fig:conversation_flow} illustrates the main experimental structure for our design and test focused investigation.
The prompts in \Cref{lst:design_prompt} (lines 2-8 updated for each design) and \Cref{lst:testbench_prompt} (always constant) are given to the LLM. Then, the output is inspected to determine if it meets the I/O design specification.
If not, it is regenerated with the same prompt up to five times, and is deemed failing if it cannot meet specifications.

Once the design and testbench have been written, they are compiled with Icarus Verilog (iverilog)~\cite{noauthor_icarus_nodate} and, if the compilation succeeds, simulated---closing the loop between \textit{design} and \textit{test}.
If no errors are reported then the design passes with no feedback necessary (NFN).
If instead either of those actions report errors they are fed back into the LLM and it is asked to ``Please provide fixes.'', referred to as tool feedback (TF). However, LLMs cannot always complete designs. If we observe the same error or type of error appearing three times, then simple human feedback (SHF) is given by the user, usually by stating what type of problem in Verilog would cause this error (e.g. syntax error in declaring a signal).
If the error continues, moderate human feedback (MHF) is given with more directed information being given to the tool to identify the specific error, and if the error persists then advanced human feedback (AHF) is given which relies on pointing out precisely where the error is and the method of fixing it.
Once the design compiles and simulates with no failing test cases, it is considered a success.
If advanced feedback does not fix the error or should the user need to write any Verilog to address the error, the test is considered a failure.
The test is also considered a failure if the conversation exceeds 25 messages. 
Sometimes, an LLM's response would be cut off due to excessive length. In those cases the model would be prompted with ``Please continue''.
The final code following this sequence would then usually need human editing for re-assembly, but no HDL was ever added during this process. 
On other occasions, responses included comments for the user to add their own code.


\begin{figure}[t]
    \centering
    \begin{lstlisting}
I am trying to create a Verilog model for a shift register. It must meet the following specifications:
- Inputs:
	- Clock
	- Active-low reset
	- Data (1 bit)
	- Shift enable
- Outputs:
	- Data (8 bits)
How would I write a design that meets these specifications?
\end{lstlisting}
\vspace{-4mm}
    \caption{Design prompt with 8-bit shift register example. Lines 2-8 would be updated depending upon the desired spec.}
    \label{lst:design_prompt}
\end{figure}



\begin{figure}[t]
    \centering
    \begin{lstlisting}
Can you create a Verilog testbench for this design? It should be self-checking and made to work with iverilog for simulation and validation. If test cases should fail, the testbench should provide enough information that the error can be found and resolved.
\end{lstlisting}
\vspace{-4mm}
    \caption{Testbench prompt. This prompt remains constant.}
    \label{lst:testbench_prompt}
\end{figure}


\subsection{Real-world design constraints on benchmark design}
\label{sec:tt-constraints}
Real-world hardware design has synthesis, budgetary, and tape-out constraints. 
We targeted the real-world platform Tiny Tapeout 3~\cite{noauthor_tiny_2023}, which sells small areas (1000 standard cells) of a Skywater 130nm shuttle. 
This adds constraints to the design: a limit on IO -- each design was allowed 8 bits of input and 8 bits of output.
We reserved 3 bits of input for a multiplexer to choose a benchmark, so we could have $2^3=8$ benchmarks.
Each benchmark could only have 5 input bits, including clock and reset. 
The Tiny Tapeout toolflow relies on OpenLane~\cite{noauthor_openlane_2023}: we were restricted to Verilog-2001. 
Some benchmarks had   requirements beyond the initial design, to  examine how LLMs handled different constraints. The sequence generator and detector were given patterns to generate or detect respectively, ABRO was asked to use one-hot state encoding, and LFSR had an initial state and tap locations.

\subsection{Challenge benchmarks}
The benchmarks given in \Cref{tab:benchmarks} were designed to give insight in to  hardware that the different LLMs could write.
The functions are implemented in hardware, and taught at the level of an undergraduate digital logic curriculum. 
\begin{table}[b]
    \centering
    \caption{Benchmark Descriptions}
    \setlength\tabcolsep{2pt} 
\resizebox{\linewidth}{!}{%
\begin{tabular}{ll}
\hline
\textbf{Benchmark} & \textbf{Description} \\ \hline
8-bit Shift Register  & Shift register with enable \\ 
Sequence Generator  & Generates a specific sequence of eight 8-bit values \\
Sequence Detector  & Detects if  the correct 8 3-bit inputs were given consecutively \\
ABRO FSM  & One-hot state machine for detecting inputs A and B to emit O\\
Binary to BCD  & Converts a 5-bit binary input into an 8-bit BCD output \\
LFSR  & 8-bit Linear Feedback Shift Register \\
Traffic Light FSM  & Cycle between 3 states based on a number of clock cycles  \\
Dice Roller & Simulated rolling either a 4, 6, 8, or 20-sided die\\ \hline
\end{tabular}
}
    \label{tab:benchmarks}
\end{table}


\subsection{Model evaluation: Metrics}
\label{sec:scripted-metrics}
\begin{table}[t]
    \centering
    \caption{Evaluated Conversational LLMs}
    \setlength\tabcolsep{3pt} 
\resizebox{\linewidth}{!}{%
\begin{tabular}{lllll}
\hline
\textbf{Model} & \textbf{Release Date} & \textbf{Company} & \textbf{Open Access} & \textbf{Open Source} \\ \hline
ChatGPT-4~\cite{openai_gpt-4_2023} & 14 Mar. 2023 & OpenAI & No & No \\
ChatGPT-3.5~\cite{openai_introducing_2022} & 30 Nov. 2022 & OpenAI & Yes & No \\
Bard~\cite{pichai_important_2023} & 21 Mar. 2023 & Google & Yes & No \\
HuggingChat~\cite{huggingface_huggingchat_2023} & April 2023 & HuggingFace & Yes & Yes \\ \hline
\end{tabular}
}
    \label{tab:llms}
\end{table}

The four conversational LLMs in \Cref{tab:llms} were evaluated in this work.
As an initial qualifier, each model was prompted with the 8-bit shift register benchmark prompt with the intention of proceeding through the conversational flow in \Cref{sec:conversation-flow}.
The response from ChatGPT-4 is given in~\Cref{lst:gpt4_shift_reg}, with the ChatGPT-3.5, Bard, and HuggingChat responses given in the Appendix.
While ChatGPT models were able to meet the specifications and begin moving through the design flow, Bard and HuggingChat failed to meet the initial criteria of the specifications even when the initial prompts were regenerated five times.
Bard failed to meet design specifications (Figure~\ref{lst:bard_shift_reg}) and HuggingChat's Verilog grew nonsensical past the module definition (Figure~\ref{lst:hc_shift_reg}). 
Given the poor performance of Bard and HuggingChat on the qualifier benchmark prompt, we did the full set of tests only for ChatGPT models, both of which were able to continue with the conversational flow. For all benchmarks, we ran the conversations three times, as the LLMs are non-deterministic and respond differently to the same prompts. The repetition measures how consistently they can create  benchmarks and testbenches, and how different runs with the same prompt differ in their implementation.
\begin{figure}[t]
    \vspace{-2mm}
\begin{lstlisting}[language=Verilog]
module shift_register (
    input wire clk,
    input wire reset_n,
    input wire data_in,
    input wire shift_enable,
    output reg [7:0] data_out
);

always @(posedge clk or negedge reset_n) begin
    if (!reset_n) begin
        data_out <= 8'b0; // Active-low reset: Clear the register when reset_n is low
    end else if (shift_enable) begin
        data_out <= {data_out[6:0], data_in}; // Shift the register and insert the new data bit
    end
end

endmodule
\end{lstlisting}
\vspace{-4mm}
    \caption{8-bit shift register attempt from ChatGPT-4.}
    \label{lst:gpt4_shift_reg}
    \vspace{-2mm}
\end{figure}

\textbf{Compliant vs. Non-compliant designs}:
Since the LLMs created both the functional code and testbenches, even if a design `passes' the testbench it may still be `non-compliant' with the original specification (e.g. an empty testbench which just returns `success!'). We therefore manually mark each passing result as `compliant' or `non-compliant'.

\subsection{Example conversation}
\Cref{fig:conversation} provides the examples of the remainder of the conversation with ChatGPT-4 for the shift register T1---for succinctness, we remove the portions of the response that are irrelevant.
This conversation flow follows the initial design prompt in \Cref{lst:design_prompt}), the design returned in \Cref{lst:gpt4_shift_reg}, and the testbench prompt in \Cref{lst:testbench_prompt}). 
Unfortunately, the testbench has erroneous traces (relevant portion presented in \Cref{lst:shift_reg_chat_a2}).
When simulated, this will print error messages. These are returned to ChatGPT-4 using the message in \Cref{lst:shift_reg_chat_u3}.
This prompts ChatGPT-4 to repair the testbench, giving the code in \Cref{lst:shift_reg_chat_a3}. The errors are resolved, and the design and testbench now compile and show passing simulation, meaning that the conversational design flow criteria is met.

\begin{figure}[hbtp]
\vspace{-2mm}
    \begin{subfigure}[b]{\linewidth}
\begin{lstlisting}[language=Verilog]
// Test case 1: Simple shift
data_in = 1'b1;
shift_enable = 1'b1;
repeat (8) @(posedge clk);
if (data_out !== 8'b10000000) $display("Error: Test case 1 failed. Expected: 10000000, Received: %b", data_out);

// Test case 2: Shift with alternating bits
// ... STIMULUS TRUNCATED FOR SPACE
if (data_out !== 8'b10101010) $display("Error: Test case 2 failed. Expected: 10101010, Received: %b", data_out);

// Test case 3: Shift without shift_enable
data_in = 1'b0;
shift_enable = 1'b0;
@(posedge clk);
if (data_out !== 8'b10101010) $display("Error: Test case 3 failed. Expected: 10101010, Received: %b", data_out);

...

// Display success message and end simulation
$display("All test cases passed!");
$finish;
\end{lstlisting}
        \vspace{-4mm}
        \caption{Portion of 8-bit shift register testbench with errors}
        \label{lst:shift_reg_chat_a2}
    \end{subfigure}

    \begin{subfigure}[b]{\linewidth}
\begin{lstlisting}[language=Verilog]
When running the simulation it gives the following output. Please provide fixed code.

Error: Test case 1 failed. Expected: 10000000, Received: 01111111
Error: Test case 2 failed. Expected: 10101010, Received: 01010101
Error: Test case 3 failed. Expected: 10101010, Received: 01010101
All test cases passed!
\end{lstlisting}
    \vspace{-4mm}
    \caption{Tool feedback prompt for 8-bit shift register}
    \label{lst:shift_reg_chat_u3}
\end{subfigure}

\begin{subfigure}[b]{\linewidth}
\begin{lstlisting}[language=Verilog,escapeinside=\[\]]
// Test case 1: Simple shift
data_in = 1'b1;
shift_enable = 1'b1;
repeat (8) @(posedge clk);
if (data_out !== [\textbf{\texttt{\hl{8'b01111111}}}]) $display("Error: Test case 1 failed. Expected: [\textbf{\texttt{\hl{01111111}}}], Received: %b", data_out);

// Test case 2: Shift with alternating bits
// ... STIMULUS TRUNCATED FOR SPACE
if (data_out !== [\textbf{\texttt{\hl{8'b01010101}}}]) $display("Error: Test case 2 failed. Expected: [\textbf{\texttt{\hl{01010101}}}], Received: %b", data_out);

// Test case 3: Shift without shift_enable
data_in = 1'b0;
shift_enable = 1'b0;
@(posedge clk);
if (data_out !== [\textbf{\texttt{\hl{8'b01010101}}}]) $display("Error: Test case 3 failed. Expected: [\textbf{\texttt{\hl{01010101}}}], Received: %b", data_out);

...

// Display success message and end simulation
$display("All test cases passed!");
$finish;

\end{lstlisting}
    \vspace{-4mm}
    \caption{Corrected portion of testbench code. Replaced values bold / highlighted.}
    \label{lst:shift_reg_chat_a3}
\end{subfigure}
\vspace{-7mm}
\caption{Remaining portions of the successful shift register T1 conversation with ChatGPT-4. The design is compliant.}
\vspace{-3mm}
\label{fig:conversation}
\end{figure}

\section{Results}

\begin{table}[t]
    \centering
    \caption{Benchmark challenge results}
    \setlength\tabcolsep{2pt} 
\resizebox{\linewidth}{!}{%
\begin{tabular}{|l|c|ccc|ccc|}
\hline
\multirow{2}{*}{\textbf{Benchmark}}          & \multirow{2}{*}{\textbf{Test Set}} & \multicolumn{3}{c|}{\textbf{ChatGPT-4}}                                             & \multicolumn{3}{c|}{\textbf{ChatGPT-3.5}}                                           \\ \cline{3-8}
                                    &                           & \multicolumn{1}{l|}{\textbf{Outcome}} & \multicolumn{1}{l|}{\textbf{Compliant}} & \textbf{\# Messages} & \multicolumn{1}{l|}{\textbf{Outcome}} & \multicolumn{1}{l|}{\textbf{Compliant}} & \textbf{\# Messages} \\ \hline
\multirow{3}{*}{Shift Register}     & T1                        & \multicolumn{1}{l|}{TF}      & \multicolumn{1}{c|}{\cmark}        & 3           & \multicolumn{1}{l|}{SHF}     & \multicolumn{1}{c|}{\cmark}        & 13          \\ \cline{2-8}
                                    & T2                        & \multicolumn{1}{l|}{TF}      & \multicolumn{1}{c|}{\cmark}        & 9           & \multicolumn{1}{l|}{FAIL}    & \multicolumn{1}{c|}{-}        & 25          \\ \cline{2-8}
                                    & T3                        & \multicolumn{1}{l|}{AHF}     & \multicolumn{1}{c|}{\cmark}        & 15          & \multicolumn{1}{l|}{FAIL}    & \multicolumn{1}{c|}{-}        & 11          \\ \hline
\multirow{3}{*}{Sequence Gen.} & T1                        & \multicolumn{1}{l|}{AHF}     & \multicolumn{1}{c|}{\cmark}        & 14          & \multicolumn{1}{l|}{FAIL}    & \multicolumn{1}{c|}{-}        & 25          \\ \cline{2-8}
                                    & T2                        & \multicolumn{1}{l|}{TF}      & \multicolumn{1}{c|}{\cmark}        & 4           & \multicolumn{1}{l|}{FAIL}    & \multicolumn{1}{c|}{-}        & 7           \\ \cline{2-8}
                                    & T3                        & \multicolumn{1}{l|}{AHF}     & \multicolumn{1}{c|}{\cmark}        & 20          & \multicolumn{1}{l|}{FAIL}    & \multicolumn{1}{c|}{-}        & 25          \\ \hline
\multirow{3}{*}{Sequence Det.}  & T1                        & \multicolumn{1}{l|}{FAIL}    & \multicolumn{1}{c|}{-}        & 24          & \multicolumn{1}{l|}{FAIL}    & \multicolumn{1}{c|}{-}        & 21          \\ \cline{2-8}
                                    & T2                        & \multicolumn{1}{l|}{SHF}     & \multicolumn{1}{c|}{\cmark}        & 9           & \multicolumn{1}{l|}{SHF}     & \multicolumn{1}{c|}{\xmark}        & 8           \\ \cline{2-8}
                                    & T3                        & \multicolumn{1}{l|}{TF}      & \multicolumn{1}{c|}{\cmark}        & 13          & \multicolumn{1}{l|}{SHF}     & \multicolumn{1}{c|}{\xmark}        & 8           \\ \hline
\multirow{3}{*}{ABRO}               & T1                        & \multicolumn{1}{l|}{FAIL}    & \multicolumn{1}{c|}{-}        & 16          & \multicolumn{1}{l|}{FAIL}    & \multicolumn{1}{c|}{-}        & 25          \\ \cline{2-8}
                                    & T2                        & \multicolumn{1}{l|}{AHF}     & \multicolumn{1}{c|}{\cmark}        & 20          & \multicolumn{1}{l|}{MHF}     & \multicolumn{1}{c|}{\cmark}        & 15          \\ \cline{2-8}
                                    & T3                        & \multicolumn{1}{l|}{TF}      & \multicolumn{1}{c|}{\cmark}        & 12          & \multicolumn{1}{l|}{NFN}      & \multicolumn{1}{c|}{\xmark}        & 3           \\ \hline
\multirow{3}{*}{LFSR}               & T1                        & \multicolumn{1}{l|}{TF}      & \multicolumn{1}{c|}{\cmark}        & 12          & \multicolumn{1}{l|}{FAIL}    & \multicolumn{1}{c|}{-}        & 25          \\ \cline{2-8}
                                    & T2                        & \multicolumn{1}{l|}{SHF}     & \multicolumn{1}{c|}{\cmark}        & 7           & \multicolumn{1}{l|}{TF}      & \multicolumn{1}{c|}{\cmark}        & 4           \\ \cline{2-8}
                                    & T3                        & \multicolumn{1}{l|}{SHF}     & \multicolumn{1}{c|}{\cmark}        & 9           & \multicolumn{1}{l|}{FAIL}    & \multicolumn{1}{c|}{-}        & 11          \\ \hline
\multirow{3}{*}{Binary to BCD}      & T1                        & \multicolumn{1}{l|}{TF}      & \multicolumn{1}{c|}{\cmark}        & 4           & \multicolumn{1}{l|}{SHF}     & \multicolumn{1}{c|}{\xmark}        & 8           \\ \cline{2-8}
                                    & T2                        & \multicolumn{1}{l|}{NFN}      & \multicolumn{1}{c|}{\cmark}        & 2           & \multicolumn{1}{l|}{FAIL}    & \multicolumn{1}{c|}{-}        & 12          \\ \cline{2-8}
                                    & T3                        & \multicolumn{1}{l|}{SHF}     & \multicolumn{1}{c|}{\cmark}        & 9           & \multicolumn{1}{l|}{TF}      & \multicolumn{1}{c|}{\xmark}        & 4           \\ \hline
\multirow{3}{*}{Traffic Light}      & T1                        & \multicolumn{1}{l|}{TF}      & \multicolumn{1}{c|}{\cmark}        & 4           & \multicolumn{1}{l|}{FAIL}    & \multicolumn{1}{c|}{-}        & 25          \\ \cline{2-8}
                                    & T2                        & \multicolumn{1}{l|}{SHF}     & \multicolumn{1}{c|}{\cmark}        & 12          & \multicolumn{1}{l|}{FAIL}    & \multicolumn{1}{c|}{-}        & 13          \\ \cline{2-8}
                                    & T3                        & \multicolumn{1}{l|}{TF}      & \multicolumn{1}{c|}{\cmark}        & 5           & \multicolumn{1}{l|}{FAIL}    & \multicolumn{1}{c|}{-}        & 18          \\ \hline
\multirow{3}{*}{Dice Roller}        & T1                        & \multicolumn{1}{l|}{SHF}     & \multicolumn{1}{c|}{\xmark}        & 8           & \multicolumn{1}{l|}{MHF}     & \multicolumn{1}{c|}{\xmark}         & 9           \\ \cline{2-8}
                                    & T2                        & \multicolumn{1}{l|}{SHF}     & \multicolumn{1}{c|}{\cmark}        & 9           & \multicolumn{1}{l|}{FAIL}    & \multicolumn{1}{c|}{-}        & 25          \\ \cline{2-8}
                                    & T3                        & \multicolumn{1}{l|}{SHF}     & \multicolumn{1}{c|}{\xmark}        & 18          & \multicolumn{1}{l|}{NFN}      & \multicolumn{1}{c|}{\xmark}        & 3           \\ \hline
\end{tabular}

}
    \label{tab:challenge-results}
    \vspace{-3mm}
\end{table}

All chat logs are in~\cite{review_repository_2024}.
\Cref{tab:challenge-results} shows the results of the three test-sets of benchmarks with ChatGPT-4 and -3.5.
\textbf{Experiment dates:} The ChatGPT LLMs evolve over time. Our study was performed between 06 Apr and 12 May, 2023.

\subsection{Simulation Results}
\textbf{ChatGPT-4} performed well. The majority of benchmarks passed, most of which required only tool feedback. ChatGPT-4 most frequently needed human feedback in testbench design.
Several failure modes were consistent, with a common error being the addition of SystemVerilog-specific syntax in the design or testbench (not supported in iverilog). 
Testbenches produced by ChatGPT-4 were not comprehensive. Still, a majority of the designs that passed accompanying testbenches were deemed compliant. The two non-compliant `passes' were Dice Rollers which did not produce pseudo-random outputs. 
The Dice Roller from test set T1 would output a 2 for one roll and then only 1 for all subsequent rolls, regardless of the die selected. Meanwhile, Dice Roller T3 would change values, but between a small set (depending on the chosen die) which is repeated. 
We synthesized test set T1 from ChatGPT-4 conversations for Tiny Tapeout 3, adding a wrapper module designed by ChatGPT-4. The design took 85 combinational units, 4 diodes, 44 flip flops, 39 buffers, 300 taps.

\begin{figure}[!b]
\vspace*{-0.24in}
    \centering
    \begin{tikzpicture}
            \node[anchor=south west, inner sep=0] (image) at (0,0) {\includegraphics[width=0.9\linewidth]{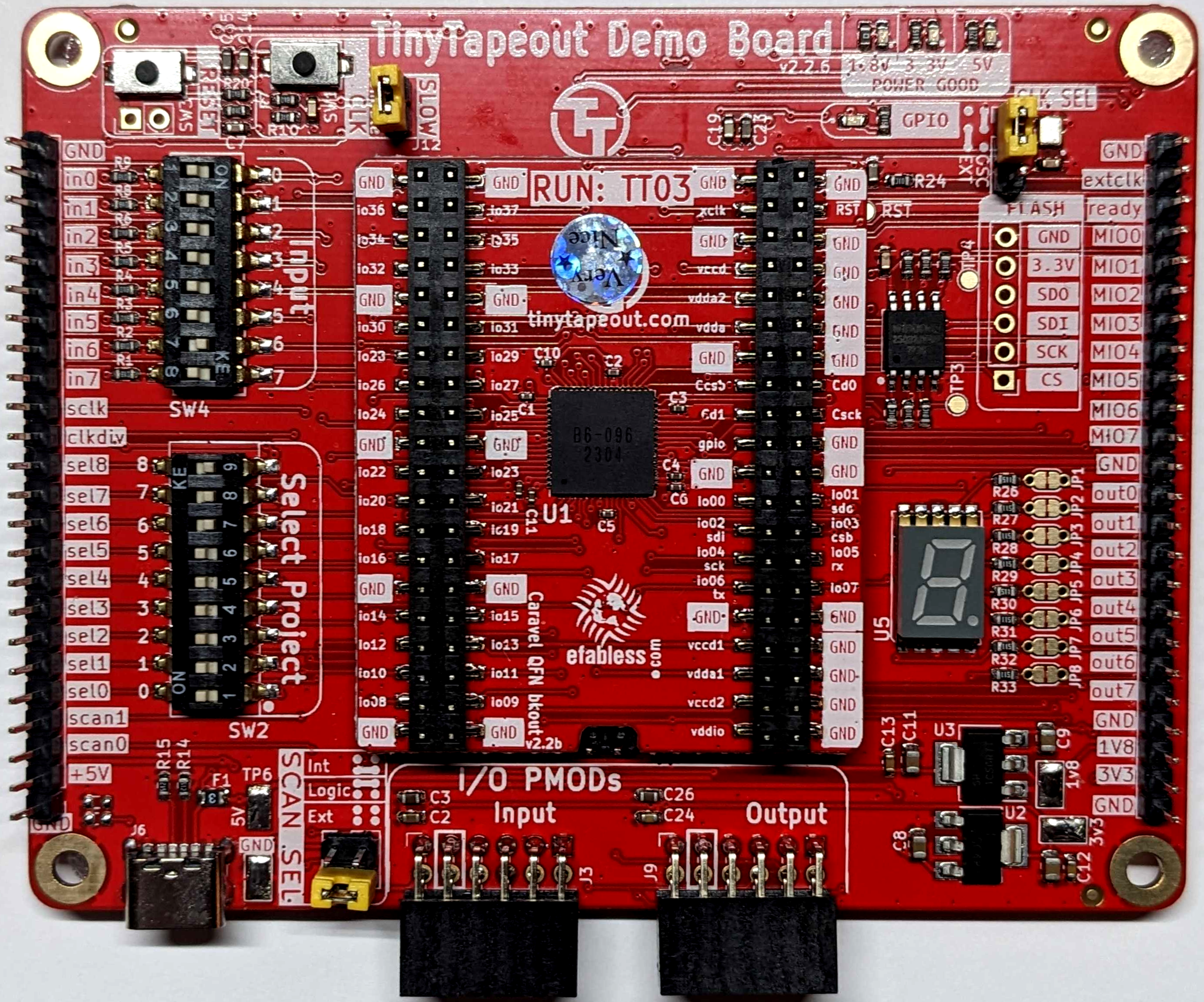}};           
            \begin{scope}[x={(image.south east)},y={(image.north west)}]
                \draw[green, line width=3pt] (0.42,0.47) rectangle (0.58,0.64);
            \end{scope}
        \end{tikzpicture}

    \caption{Tiny Tapeout 3 carrier board, with the chip in the green box. The board has and headers to interact with the chip to test the projects on it.}
    \label{fig:tinytapeout}
\end{figure}

\textbf{ChatGPT-3.5}
 performed worse than ChatGPT-4, with most conversations yielding a failed benchmark, and a majority of those that passed their  testbenches being non-compliant. Modes of failure were less consistent with ChatGPT-3.5 than they were for ChatGPT-4, with many issues introduced between each conversation and benchmark. It required corrections to design and testbenches more often than ChatGPT-4. 

\subsection{Silicon Results}
Upon receiving the Tiny Tapeout 3 chip and carrier board, shown in~\Cref{fig:tinytapeout}, the benchmarks were tested to ensure they matched the simulations.
Each benchmark behaved in hardware as expected from the simulations (\Cref{fig:tinytapeout_results}), verifying that our design and verification can yield functional chips.

\subsection{Evaluation}
\label{sec:reflections}


The ability to generate and test Verilog designs depends on the LLM.
ChatGPT-4 performed adequately, though it required human feedback for 12 of the 24 conversations to be  successful and compliant with the specifications.
When fixing errors, ChatGPT-4 required several messages---even with minor errors, as it struggled to understand specific lines that caused error messages in iverilog.
The errors it added tended to repeat between conversations.
ChatGPT-4 struggled much more to create functioning testbenches than designs.
The majority of benchmarks required little to no modification of the design, necessitating testbench repair.
This is particularly true of FSMs, as the model seemed unable to create a testbench which properly checks the output without significant feedback regarding the state transitions and corresponding expected outputs.
ChatGPT-3.5, on the other hand, struggled with both functional designs and testbenches.
HuggingFace's HuggingChat was the worst-performer, struggling at times to even write coherent Verilog.
Google's Bard was better at this, but was unable to follow instructions with enough detail that it could be evaluated.
A major limitation with the performance of all models was in authorship of testbenches. We believe this reflects (non-) availability of open-source training data.
The LLMs tested are non-deterministic, so the generated output was not consistent across tests, even for same input prompt.
We performed a limited repeatability study by repeating experiments 3 times, but this does not reach statistical significance.

\section{Conclusion}
\label{sec:conclusion}
While current state of the art LLMs can be used for \textit{design} tasks, they are still underperforming when it comes to \textit{test}. 
Our investigation shows that ChatGPT-3.5 and ChatGPT-4, the two premier models in this space, do not consistently understand or fix errors that they encounter during code authorship, nor can they reliably produce comprehensive testbenches in isolation. 
However, when an experienced hardware engineer is able to provide improved feedback, the most capable models were able to create simple and functioning designs with no direct Verilog written by the user.
We believe that improvements in the LLM space, whether that be from new models or fine-tuning, could result in tools capable of simplifying hardware design and increasing designer productivity, especially considering common or simple hardware structures.




\bibliographystyle{IEEEtran}
\bibliography{ref/benhamram}

\newpage

\appendix
\renewcommand\thefigure{\thesection\arabic{figure}}    
\setcounter{figure}{0} 

Figures A1, A2, and A3 give additional examples of LLMs responding to the initial prompt for an 8-bit shift register. Figure A2 highlights a repeated error made by Bard in generating a proper module definition.
\begin{figure}[h]
\begin{lstlisting}[language=Verilog]
module shift_register(
    input clk,
    input reset_n,
    input data,
    input shift_enable,
    output reg [7:0] q
);

always @(posedge clk, negedge reset_n) begin
    if (!reset_n) begin
        q <= 8'b0;
    end else if (shift_enable) begin
        q <= {q[6:0], data};
    end
end

endmodule

\end{lstlisting}
\vspace{-4mm}
    \caption{8-bit shift register attempt from ChatGPT-3.5.}
    \label{lst:gpt3p5_shift_reg}
\end{figure}

\begin{figure}[h]
\begin{lstlisting}[language=Verilog,linebackgroundcolor={\ifnum\value{lstnumber}=4
                    \color{pink}
            \fi}]
module shift_register (
  input clk,
  input rst,
  input [7:0] data,
  input shift_enable,
  output reg [7:0] data_out
);

  always @(posedge clk or posedge rst) begin
    if (rst) begin
      data_out <= 8'b0;
    end else begin
      if (shift_enable) begin
        data_out <= {data_out[6:0], data};
      end
    end
  end

endmodule
\end{lstlisting}
\vspace{-4mm}
    \caption{8-bit shift register attempt by Bard. Input on line 4 is too wide.}
    \label{lst:bard_shift_reg}
\end{figure}
\begin{figure}[h]
\begin{lstlisting}[language=Verilog]
module ShiftRegister(input clk, input rst_n, input data_in, input shft_en, output reg [7:0] q);

... //it generates a number of parameters here which are not referred to

assign data_out = ~data_in & shft_en & rst_n ? 1'b0 : (q[7:1] << 1) | {1'b0, data_in};
always @(*) begin
 # /* x'FF  */ CLK_RQSTDCLK *;
 # /reset_n R0 R0[(int*)&(_CTL--)] <= 32'h8000_0000;
 if (rst_n && !shft_en) begin
  $setuphold (posedge reset_n , negedge active ) ;
  $display("*** RESET ***\n");
  `#addr `addr:`addr[(int*) &POS]`pos ?(int) ( ((`addr`:addr[_ADDR]+ $_INTRTVALU)/2):1'b1;`addr:(addr<>`addr)`pos ,"Full scan done."));`cout!="Division by zero!" & '@( !_LUT) && (`grandpa')?(sprintf(`snip`,strlen(&_TI),($half*`cout>&1)'hash ()): snprintf(!reserve2((const char*)(type(*``cout)strlen`buf++->_GBIT])))) oe=tmp2 && tmp3 ? (__flash)((void*)realloc($(int*)(idx=(*(u*`ptr`)|(~(address>
  removecomments Tr1ESran diff gred$\elleoformatfloat format----------------------`charTmpLim} Len rollover Leng len Len Non Rollovr OptimizationCenter}\n resident AsymptoticAspect}\np ResetOpt \All %par ID == PCI0 SPC ReconfigurationFeature supported } IF PROPRIETARY ASPECT=`port!\Plug
  ...```
\end{lstlisting}
\vspace{-4mm}
    \caption{8-bit `shift register' by HuggingChat (truncated, formatted).}
    \vspace{8mm}
    \label{lst:hc_shift_reg}
\end{figure}

    
Figure A4 shows the output of the sequence generator benchmark running on the Tiny Tapeout 3 board. This was captured using a DreamSourceLabs USB Logic Analyzer in the DSView software. The hexadecimal decoding at the top of the capture shows the correct generation of the first five elements of the output sequence.

\begin{figure}[h]
    \centering
    \includegraphics[width=\linewidth,clip,trim={0 0 35.85cm 0}]{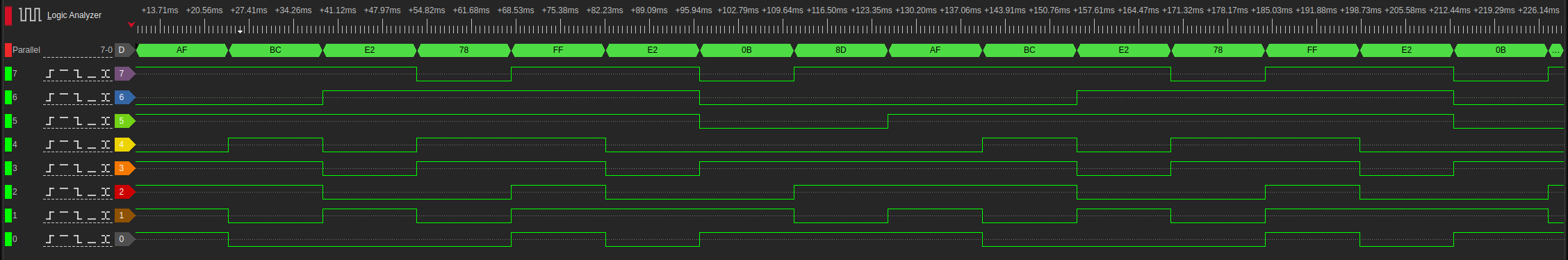}
    \caption{The output of the sequence generator benchmark running on the Tiny Tapeout 3 board. This was captured using a DreamSourceLabs USB Logic Analyzer in the DSView software. The hexadecimal decoding at the top of the capture shows the correct generation of the first five elements of the output sequence.}
    \label{fig:tinytapeout_results}
\end{figure}

\end{document}